\documentclass[12pt]{article}
\usepackage{epsf}  
\begin{document}

\newcommand{\sheptitle}
{Radiative generation of  $\bigtriangleup m^2_{21}$ and $\theta_{13}$ in two-fold degenerate neutrino models}
\newcommand{\shepauthor}
{ N. Nimai Singh$^{a,b}${\footnote{Regular Associate of ICTP; {\it{e-mail address:}} nimai03@yahoo.com}} and Mrinal K. Das$^b$ }

\newcommand{\shepaddress}
{$^a$International Centre for Theoretical Physics, Strada Costiera 11,\\ 31014 Trieste, Italy \\ and \\ 
$^b$Department of Physics, Gauhati University, Guwahati-781014, India}
\newcommand{\shepabstract}
{We study  the implications  on an important result  on radiative corrections (JM conjecture)  which 
 states that  the solar mass scale $\bigtriangleup m^2_{21}$
 corresponding to the large mixing angle  (LMA) MSW solution of the solar neutrino problem,  can  
 be generated through radiative corrections in the minimal supersymmetric standard model (MSSM) 
 if two of the neutrino masses  having opposite CP parity, are assumed to be 
degenerate and have non-zero value of  $U_{e3}$  at high energy scale. We show  that the above conjecture can be derived  when we take  the    
static solution of  mixing angles from the general solution of RGEs. 
If we consider  the simultaneous  running of the three neutrino mass eigenvalues  and mixing angles,
 the above  two-fold degenerate case $m_i=(m,-m,m')$
with  $m'\neq 0$ and $U_{e3}=0$, can  also generate through radiative corrections, the  non-zero values of  $\bigtriangleup m^2_{21}$  and $U_{e3}$ at low scale, 
 which are consistent with the LMA MSW solution. Such  generalisation of JM conjecture  may have 
 important implications for model buildings from  general gauge symmetry. }
\begin{titlepage}
\begin{flushright}
hep-ph/0407206
\end{flushright}
\begin{center}
{\large{\bf\sheptitle}}
\bigskip\\
\shepauthor
\\
\mbox{}\\
{\it\shepaddress}\\
\vspace{.5in}
{\bf Abstract}
\bigskip
\end{center}
\setcounter{page}{0}
\shepabstract
\end{titlepage}
\section{Introduction}

 In order to have a meaningful comparison of the theoretical
 predictions on neutrino masses and mixing angles at high energy scale    
 with the  low energy neutrino oscillation experiments[1,2], the effects of radiative corrections are very important[3-10].
 The general   motivations for  studying  the renormalisation group equations (RGEs) are - 
to check the stability of the neutrino mass model under   radiative corrections[7,9];
to generate solar mass scale $\bigtriangleup m^2_{21}$  and also  reactor mixing angle $|U_{e3}|$ from radiative corrections[10,11];
to generate  correct radiative magnifications of solar and atmospheric mixing angles from CKM-like small values at high scale[12]; and 
to get  suitable deviations  from the bimaximal solar and atmospheric mixings through radiative corrections [13,14], etc.
Within the framework of running the RGEs from high energy scale to low energy scale, there are  again two different  approaches so far
 employed in the literature. 
In the first approach  the running is carried out through the neutrino mass matrix $m_{LL}$ as a whole, 
and at every point in the energy scale 
one can extract neutrino masses and mixing angles through the diagonalisation of the neutrino mass matrix calculated  at that  particular energy  scale[7]. 
In the second approach  the running of the RGEs is carried out directly in terms neutrino mass eigenvalues and three mixing angles [15,16,17].

In the present paper we follow  the second approach and carry out a detailed numerical analysis of the RGEs in the minimal 
supersymmetric standard model (MSSM). We follow the RGEs  for three generations of neutrino masses and mixings from Ref.[15].     
 We are particularly interested to study a profound  result on the effect of radiative corrections ( we call it as  JM conjecture) [11]
 which states that 
electroweak radiative correction in MSSM can   generate the solar neutrino mass scale  $\bigtriangleup m^2_{21}$ required for the LMA MSW 
solution to the solar neutrino problem if two
of the neutrino masses  with opposite CP parity, are assumed to be  degenerate at high energy scale  and the input high energy  value of $U_{e3}$ 
is non-zero.
 We show from the general RGEs that  the  above JM result[11] is limited to 
static approximation of mixing angles, and hence it is expected to have  limited predictions. We generalise the JM result [11]  by including 
the evolution of mixing angles and present the numerical results which have profound 
implications in model buildings on neutrino masses and mixings from general gauge theories.

\pagebreak

\section{Formalism}

The left-handed Majorana neutrino mass matrix $m_{LL}$ which is generally obtained from seesaw mechanism at high scale $M_R$, 
is usually  expressed in terms of $K(t)$, the coefficient of the dimension five neutrino mass operator[3,4] in a scale-dependent manner,
 $t=\ln(\mu/1GeV)$ [7],
\begin{equation}
m_{LL}(t)=v^2_{u}K(t)
\end{equation}
where the vacuum expectation value (vev) is $v_{u}=v_0 \sin\beta$ and $v_0=174$ GeV in MSSM.
The neutrino  mass eigenvalues $m_i$  and the PMNS mixing matrix $U_{PMNS}$ [18] are then extracted  through the diagonalisation of $m_{LL}(t)$ at every point 
in the energy scale $t$,
\begin{equation}
m_{LL}^{diag}=Diag(m_{1},m_{2},m_{3})=V_{\nu L}m_{LL}V_{\nu L}^{T}
\end{equation}
and the PMNS mixing matrix $U_{PMNS}=V^{\dag}_{\nu L}$ where  
\begin{equation}
U_{PMNS}=\left(\begin{array}{ccc}
U_{e1} & U_{e2} & U_{e3} \\
U_{\mu 1} & U_{\mu 2} & U_{\mu 3} \\
U_{\tau 1} & U_{\tau 2} & U_{\tau 3}
\end{array}\right)
\end{equation}
 is usually  parametrised in terms  of the product of three rotations  $R(\theta_{23})$, $R(\theta_{13})$ and  $R(\theta_{12})$,
 (neglecting CP violating phases) by 
\begin{equation}
U_{PMNS}=\left(\begin{array}{ccc}
c_{13}c_{12}  & c_{13}s_{12}  & s_{13} \\
-c_{23}s_{12}-c_{12}s_{13}s_{23}  & c_{12}c_{23}-s_{12}s_{13}s_{23} & c_{13}s_{23} \\
s_{12}s_{23}-c_{12}s_{13}c_{23} & -c_{12}s_{23}-c_{23}s_{13}s_{12} & c_{13}c_{23}
\end{array}\right)
\end{equation}
where $s_{ij}=\sin{\theta_{ij}}$ and $c_{ij}=\cos{\theta_{ij}}$  respectively. 
Here the three mixing angles are  defined as  $\tan\theta_{12}=|U_{e2}|/|U_{e1}|$,  $\tan\theta_{23}=|U_{\mu 3}|/|U_{\tau 3}|$ and $\sin\theta_{13}=|U_{e3}|$.
The solar LMA data favours the 'light-side', usually denoted by   $\tan\theta_{12}<1$ [19]
 for the usual sign convention $|m_2|>|m_1|$. We restrict our future  analysis within this convention only and any interchange of first two mass eigenvalues 
amounts to interchanging the ordering of the 1st two columns on $U_{PMNS}$.
 There are also certain physical  arguments in favour of  presenting data in terms of $\sin^2\theta$, where 
$\theta$ is the mixing angle appropriate for a given experiment[20].


The  RGEs for the   eigenvalues of coefficient $K(t)$ in eq.(1), are expressible as [15]
\begin{equation}
\frac{d}{dt}K_{i}=\frac{1}{16\pi^2}\sum_{f=e,\mu,\tau}[(-\frac{6}{5}g_{1}^{2}-6g_{2}^2+6Tr(h_{u}^2)+2h_{f}^{2}U_{fi}^{2}]K_{i}
\end{equation}
Neglecting $h_{\mu}^2$ and $h_{e}^2$ compared to $h^2_{\tau}$, and taking scale-independent vev as in eq.(1), we have the complete RGEs
 for three neutrino mass eigenvalues, 
\begin{equation}
\frac{d}{dt}m_{i}=\frac{1}{16\pi^2}[(-\frac{6}{5}g_{1}^{2}-6g_{2}^2+6h_{t}^2)+2h_{\tau}^{2}U_{\tau i}^{2}]m_{i}
\end{equation}

The  approximate analytical solution of the above equation can be obtained  by taking static  mixing angle $U_{\tau i}^2$ in the integration range, as [16]
\begin{equation}
m_{i}(t_{0})=m_{i}(t_R)exp(\frac {6}{5} I_{g1}+6I_{g2}-6I_{t})exp(-2U_{\tau i}^2 I_{\tau})
\end{equation} 
The integrals in the above expression  are usually  defined as[6,7] 
\begin{equation}
I_{gi}(t_{0})=\frac{1}{16\pi^2}\int_{t_{0}}^{t_{R}}g_{i}^{2}(t)dt
\end{equation}
and,
\begin{equation}
I_{f}(t_{0})=\frac{1}{16\pi^2}\int_{t_{0}}^{t_{R}}h_{f}^{2}(t)dt
\end{equation}
where $i=1,2,3$ and $f=t,b,\tau$ respectively.
For a two-fold degenerate neutrino masses  with opposite CP parity at high scale, viz.,
 $m_{LL}^{diag}= Diag.(m, -m, m')=U_{PMNS}^T m_{LL}U_{PMNS}$, 
 eq.(7) is further  simplified to the following expressions
\begin{equation}
m_{1}(t_0)\approx m(t_R)(1+2\delta_{\tau}(c_{12}s_{13}c_{23}-s_{12}s_{23})^2)+O(\delta_{\tau}^2), 
\end{equation}
\begin{equation}
m_{2}(t_0)\approx -m(t_R)(1+2\delta_{\tau}(c_{23}s_{13}s_{12}+c_{12}s_{23})^2)+O(\delta_{\tau}^2), 
\end{equation}
\begin{equation}
m_{3}(t_0)\approx m'(t_R)(1+2\delta_{\tau}(c_{13}c_{23})^2)+O(\delta_{\tau}^2).
\end{equation} 
While deriving the above expressions, the following approximations are made,
$$ \exp(-2|U_{\tau i}|^2 I_{\tau})\simeq 1-2|U_{\tau i}|^2 I_{\tau}=1+2|U_{\tau i}|^2 \delta_{\tau},$$
$$ -\delta_{\tau}=I_{\tau} \simeq \frac{1}{\cos ^2\beta}(m_{\tau}/4 \pi v)^2 \ln (M_R/m_t),$$
$$ f=\exp(\frac{6}{5}I_{g1}+6I_{g2}-6I_{t})\approx 1.$$
It is clear that  $\delta_{\tau}$ is a negative quantity in MSSM and the low energy solar neutrino mass scale is then obtained as 
\begin{equation}
\bigtriangleup m^2_{21}(t_0)= m^2_{2}-m^2_{1}\approx 
4 \delta_{\tau}m^2(\cos 2\theta_{12}(s^2_{23}-s^2_{13}c^2_{23})+s_{13}\sin 2\theta_{12}\sin 2\theta_{23})+O(\delta^2_{\tau})
\end{equation}
For  certain limited range of  $s_{13}$, the  second term on the RHS in the above expression (13)  may lead to positive value of  $\bigtriangleup m^2_{21}$ 
consistent with the LMA solution  for $\tan\theta_{12} < 1$. This is the main result of the  JM conjecture [11] derived from a completely  different approach.
 But as we see from eq.(7), it is based on the approximation of static  mixing angles between the two energy scales. 
This point will be addressed  in the present numerical  analysis 
by adding the running of the mixing angles along with the evolution of  neutrino masses.   

The corresponding evolution equations for the PMNS matrix elements $U_{fi}$ are given by [15]
\begin{equation}
\frac{d U_{fi}}{dt}=-\frac{1}{16\pi^2}\sum_{k\neq i}\frac{m_{k}+m_{i}}{m_{k}-m_{i}} U_{fk}(U^TH^2_eU)_{ki}
\end{equation}
where  $f=e,\mu,\tau$ and $i=1,2,3$ respectively. Neglecting $h_{\mu}^2$ and $h_{e}^2$ as before, and denoting $A_{ki}=\frac{m_{k}+m_{i}}{m_{k}-m_{i}}$, 
eq.(14) simplifies to [15]
\begin{equation}
\frac{d s_{12}}{dt}=\frac{1}{16\pi^2}h_{\tau}^2 c_{12}[c_{23}s_{13}s_{12}U_{\tau 1}A_{31}-c_{23}s_{13}c_{13}U_{\tau 2}A_{32}
+U_{\tau 1}U_{\tau 2} A_{21}],
\end{equation}

\begin{equation}
\frac{d s_{13}}{dt}=\frac{1}{16\pi^2}h_{\tau}^2 c_{23}c_{13}^2[c_{12}U_{\tau 1}A_{31}+s_{12}U_{\tau 2}A_{32}],
\end{equation}

\begin{equation}
\frac{d s_{23}}{dt}=\frac{1}{16\pi^2}h_{\tau}^2 c_{23}^2[-s_{12}U_{\tau 1}A_{31}+c_{12}U_{\tau 2}A_{32}].
\end{equation}
For the two-fold degenerate case with opposite CP parity under study, $m_i=(m, -m, m')$,   eq.(16) can be further simplified  for two cases:\\
Case (i): For input values $m_3=m'=0$, $s_{13}\neq 0$, 
\begin{equation}
\frac{d s_{13}}{dt}=\frac{1}{16\pi^2}h_{\tau}^2 c_{23}c_{13}^2[c_{23}s_{13}];
\end{equation}

Case (ii): For input values $m_3=m'\neq 0$, $s_{13}= 0$, 
\begin{equation}
\frac{d s_{13}}{dt}=\frac{1}{16\pi^2}h_{\tau}^2 c_{23}c_{13}^2[c_{12}s_{12}s_{23}(A_{31}-A_{32})].
\end{equation}
In  both cases the mixing angle $s_{13}$ evolves with energy scale, and this  in turn induces solar mass scale through eq.(13).
However, if both $m_3=m'=0$ and $s_{13}=0$, the RHS of eq.(16) reduces to zero, and 
hence  the angle $s_{13}$ ceases to  evolve further with energy, and hence   no generation of solar scale. 
 The above point in case (ii) is a generalisation to  JM result[11]. However for numerical analysis we confine to the original RG equations (7), (15)- (17)
 for neutrino masses and mixings.

\pagebreak

\section{Numerical analysis  and results}
For a complete numerical analysis of the RGEs, eqs. (7),(15)-(17) outlined in the previous section, 
we follow  here two consecutive steps (a)   bottom-up running[6]  in the first one, and (b)  top-down running[7] in the next.
 In the first step (a),  the running 
of the RGEs for the third family Yukawa couplings $(h_{t},h_{b}, h_{\tau})$ and three gauge couplings $(g_{1},g_{2},g_{3})$ 
in MSSM , are carried out  from top-quark mass scale ($t_0$) at low energy end to high energy scale $M_{R}$ where $B-L$ symmetry breaks down[6,16].
In the present analysis we consider the high scale  $M_{R}=10^{13}$GeV as well as the unification scale 
 $M_R=1.6\times 10^{16}GeV$ and take the large $\tan\beta$ input value  $\tan\beta=55, 58, 60$. 
Normally SUSY breaking scale is taken around 1 TeV, but for simplicity of the calculation we assume here  at the top-quark 
mass scale $t_0=\ln m_t$[6,7].
We adopt the  standard procedure to get the  values of gauge couplings at top-quark mass scale from the experimental CERN-LEP measurements 
at $M_{Z}$, using one-loop RGEs, assuming the existence of a 
one-light Higgs doublet and five quark flavours below $m_t$ scale[6,16]. Similarly, the  Yukawa couplings are also evaluated at top-quark mass 
scale using QCD-QED 
rescaling factors in the standard fashion[16]. 
 In the second stage (b), the  running of
 three neutrino masses $(m_1, m_2, m_3)$  and mixing angles  $(s_{12}, s_{23}, s_{23})$ is carried out together with the 
running of the  Yukawa and gauge cuoplings,
 from the high scale $t_R$ to low scale $t_0$. In this case we use the values of Yukawa and gauge couplings evaluated earlier 
 at the scale $t_R$ from the first stage runnung of RGEs in (a).
In principle one can evaluate neutrino masses and mixing angles at every point in the energy scale.

As discussed in the previous section, we study here the two-fold degenerate neutrino mass model with opposite CP parity, $m_i=(m, -m, m')$
at high scale.
Radiative generations  of both $\bigtriangleup m^2_{21}$ and $\sin\theta_{13}$ at low energies can be studies under three 
different cases  at high scale: (a) when $m'\neq 0$ and $s_{13}\neq 0$; (b) when $m'=0$ and $s_{13}\neq 0$ ; (c) when 
$m'=0$ and $s_{13}= 0$; and  (d) when $m'\neq 0$ and $s_{13}=0$.
Cases (a) to  (c)  have  been covered   in the analysis of  JM paper [11] and showed consistency within MSSM electroweak corrections.
We are concerned here with case (d) and its implications as a generalisation to JM conjecture. This was not possible in  the JM analysis 
 as the equations  were  derived under the assumption on  static mixing angles. When we use both 
runnings  of mass egenvalues and mixing angles in the RGEs, the non-zero input  value of $m'$ will first induce a  non-zero values of $s_{13}$ 
which in turn generates non-zero value of $\bigtriangleup m^2_{21}$ while running from high to low scale.

 With the recent  analysis[2] of the new KamLAND spectral data, the solar mass scale has more precise best-fit value 
$\bigtriangleup m^2_{21}=8.4\times 10^{-5}eV^2$ and solar angle $\sin\theta_{12}=0.24$. The corresponding values at $3\sigma$-level  are 
$\bigtriangleup m^2_{21}=(7.2 - 9.5)\times 10^{-5} eV^2$ and $\sin^2\theta_{12}=0.21 - 0.37$. However the bound on $\sin^2\theta_{13}$ from the combined 
Solar, CHOOZ and KamLAND, leads to $\sin^2\theta_{13}<0.05$ [2,21]. 
We present below our results irrespective of these  experimental data, though 
there are enough scope for fine tunning of our predictions through choices of input parameters.

We present the results in Tables 1-4 where the parameters $m^0_1$, $m^0_2$, $m^0_3$, $s^0_{12}$, $s^0_{23}$, $s^0_{13})$ are 
the  arbitrary input values at the high energy scale. The other parameters in the lower half of the tables represent the values 
evaluated at low scale through radiative corrections. In order to get more effects of radiative corrections we always take larger 
$\tan\beta$ values in the calculations. We also consider two separate high scale values $M_R=10^{13}$GeV and 
$M_R=M_U=1.6\times 10^{16}$GeV. In Tables 1-3 we study the case (d) which has input values $m'\neq 0$ and $s_{13}=0$  for two  different high
 scales $10^{13}GeV$ and $1.6\times 10^{16}GeV$ and also for different values of $\tan\beta= 55, 58, 60$; while for cases (a) to (c),
 the numerical results are presented 
in Table 4 for completeness[17].   Table 3 is particularly interesting 
as we have taken common input values at high scale $(M_R=1.6\times 10^{16}GeV)$  for different $\tan\beta$ values.  We observe better results 
for  higher $\tan\beta$ values.
We have checked that for case where both $m'=0$ and $U_{e3}=0$ at high scale , it is not possible to generate solar mass  scale at low energy 
 in MSSM ( Table 4). This is consistent with the theoretical prediction in the previous section.  In short, the results in 
 Table 4 agree  well with JM conjecture[11]
 and the results in Tables 1-3 are the generalisation to JM results,  when the runnings of mixing angles are taken into account.

 For input parameters $\tan\beta=60$ and  $M_R=1.6\times 10^{16}GeV$,  we present 
in Figs. 1 and 2 the generation of 
CHOOZ mixing angle $\sin\theta_{13}$ and $\bigtriangleup m^2_{21}$ respectively. The solid-lines in the graphs are  based on the input parameters  given 
in Table-3 for $\tan\beta=60$. Here we have $s^0_{23}=s^0_{12}=0.707107$ as input values and  the solar   scale $\bigtriangleup m^2_{21}$  increases  with 
the decrease in energy scale (Fig.2).
  The input parameters  for the dashed-lines and dotted-lines in the two graphs (Figs.1,2) are again taken from 
last two columns (4th and 5th) in Table-1 respectively. Here  the value of either $s^0_{12}$ or  $s^0_{23}$ is  deviated from $0.707107$.
The changes in input angles make  the evolution pattern in solar 
mass scale  quite different and irregular (for dashed and dotted lines) as seen in Fig.2.
However the evolution pattern  in CHOOZ angle is uniform  over the wide variations of input values of solar and atmospheric 
mixing angles as seen in Fig.1. Thus it is possible to generate simultaneously both solar mass scale and CHOOZ mixing  angle through radiative corrections 
in the present two-fold degenerate model.\\

Table 1:Radiative generation of $\bigtriangleup m_{21}^2$ and $U_{e3}$ in MSSM while running from high scale  $M_R=1.6\times10^{16}$GeV
 to low scale $m_t=175GeV$ for different input values of $\tan\beta$.
 The parameters $(m^0_{1,2,3},s^0_{23,13,12})$ are defined at high scale while other predictions are at low scale.

\begin{tabular}{lllll} \\ \hline
Parameter & $\tan\beta=55$ & $\tan\beta=58$ & $\tan\beta=60$ & $\tan\beta=60$ \\ \hline              
$m_{1}^0[eV]$ & 0.0973 & 0.0975 & 0.096   & 0.09 \\
$m_{2}^0[eV]$ & -0.0973 & -0.0975 & -0.096 & -0.09\\
$m_{3}^0[eV]$ & -0.088  & -0.083  & -0.079  & -0.07\\
$s_{23}^0 $   & 0.82 &    0.81    & 0.78    & 0.707107 \\           
$s_{13}^0 $   & 0.00  &   0.0     & 0.0     & 0.0 \\           
$s_{12}^0 $   & 0.63  &   0.635   & 0.635   & 0.65 \\ \hline           
$m_{1}[eV] $      & 0.068350 & 0.064487 & 0.060151 & 0.056928\\           
$m_{2}[eV] $      & -0.068894 & -0.064920 & -0.060801 & -0.057506\\           
$m_{3}[eV] $      & -0.059309 & -0.052094 & -0.045483  & -0.039571\\           
$s_{23} $     & 0.5493 & 0.5804 & 0.5450               &  0.5237 \\           
$s_{13}$      & -0.2860 & -0.2495 & -0.2422           & -0.1854 \\
$s_{12} $     & 0.5853  & 0.6022 & 0.6038                & 0.63295 \\             
$\bigtriangleup m_{12}^2[10^{-5}eV^2]$ & 7.462 & 5.605 & 6.505 & 6.605 \\         
$\bigtriangleup m_{23}^2[10^{-3}eV^2]$ & 1.229 & 1.50 & 1.66 & 1.746\\ \hline         
\end{tabular}

\pagebreak

Table 2: Radiative generation of $\bigtriangleup m_{21}^2$ and $U_{e3}$ in MSSM while running from high scale  $M_R=10^{13}$GeV
 to low scale $m_t=175GeV$ for different input values of $\tan\beta$.
 The parameters $(m^0_{1,2,3},s^0_{23,13,12})$ are defined at high scale while other predictions are at low scale.

\begin{tabular}{llll} \\ \hline
Parameter & $\tan\beta=55$ & $\tan\beta=58$ & $\tan\beta=60$ \\ \hline              
$m_{1}^0[eV]$ & 0.0974 & 0.0974 & 0.096 \\
$m_{2}^0[eV]$ & -0.0974 & -0.0974 & -0.096\\
$m_{3}^0[eV]$ & -0.09  & -0.088  & -0.085 \\
$s_{23}^0 $   & 0.83 &    0.82    & 0.83  \\           
$s_{13}^0 $   & 0.00  &   0.0     & 0.0   \\           
$s_{12}^0 $   & 0.643  &   0.64   & 0.64  \\ \hline           
$m_{1}[eV] $      & 0.071131 & 0.069098 & 0.066296 \\           
$m_{2}[eV] $      & -0.071599 &-0.069583 &-0.066779 \\           
$m_{3}[eV] $      & -0.064140 & -0.060376 & -0.056698 \\           
$s_{23} $     & 0.58158 & 0.57746 & 0.59112 \\           
$s_{13}$      & -0.28135 & -0.2698 & -0.26964 \\           
$s_{12} $     & 0.60227  & 0.60233 & 0.60225 \\             
$\bigtriangleup m_{12}^2[10^{-5}eV^2]$ & 6.666 & 6.735 & 6.433 \\         
$\bigtriangleup m_{23}^2[10^{-3}eV^2]$ & 1.012 & 1.197 & 1.245\\ \hline         
\end{tabular}
\\

\pagebreak

Table 3:Radiative generation of $\bigtriangleup m_{21}^2$ and $U_{e3}$ in MSSM while running from high scale  $M_R=1.6\times10^{16}$GeV
 to low scale $m_t=175GeV$ for different input values of $\tan\beta$.
 The high energy parameters $(m^0_{1,2,3},s^0_{23,13,12})$ are common for all three cases of input $\tan\beta$ values.

\begin{tabular}{llll} \\ \hline
Parameter & $\tan\beta=60$ & $\tan\beta=58$ & $\tan\beta=55$\\ \hline              
$m_{1}^0[eV]$ & 0.09 & 0.09 & 0.09 \\
$m_{2}^0[eV]$ & -0.09 & -0.09 & -0.09\\
$m_{3}^0[eV]$ & -0.05  & -0.05  & -0.05 \\
$s_{23}^0 $   & 0.707107 &    0.707107    & 0.707107  \\           
$s_{13}^0 $   & 0.00  &   0.0     & 0.0   \\           
$s_{12}^0 $   & 0.707107  &   0.707107  & 0.707107  \\ \hline           
$m_{1}[eV] $      & 0.056374 & 0.059756 & 0.063428 \\           
$m_{2}[eV] $      & -0.057110 &-0.060265 & -0.063752 \\           
$m_{3}[eV] $      & -0.028741 & -0.031086 & -0.033575 \\           
$s_{23} $     & 0.620455 & 0.637142 & 0.652951 \\           
$s_{13}$      & -0.09362 & -0.07728 & -0.06103 \\           
$s_{12} $     & 0.7032  & 0.70449 & 0.70550 \\             
$\bigtriangleup m_{12}^2[10^{-5}eV^2]$ & 8.3587 & 6.112 & 4.129 \\         
$\bigtriangleup m_{23}^2[10^{-3}eV^2]$ & 2.4356 & 2.6655 & 2.937\\ \hline         
\end{tabular}

\pagebreak

Table 4: Analysis of JM conjecture related to radiative generation of $\bigtriangleup m_{21}^2$ and $U_{e3}$ in MSSM 
while running from high scale  $M_R=10^{13}$GeV
 to low scale $m_t=175GeV$ for different cases. The  parameters $(m^0_{1,2,3},s^0_{23,13,12})$ are defined at high  energy scale 
while others are at low energy. Meanings of Cases (a,b,c) are explained in the text. 

\begin{tabular}{llll} \\ \hline
Parameter & Case (a)           & Case (b)                  & Case (c)\\ \hline              
$m_{1}^0[eV]$ & 0.08           & 0.095                     & 0.08 \\
$m_{2}^0[eV]$ & -0.08          & -0.095                    & -0.08\\
$m_{3}^0[eV]$ & -0.04          &  0.0                      & 0.0 \\
$s_{23}^0 $   & 0.707107       & 0.707107                  & 0.707107  \\           
$s_{13}^0 $   & 0.1            &  0.1                      & 0.0   \\           
$s_{12}^0 $   & 0.707107       &  0.707107                 & 0.707107  \\ \hline           
$m_{1}[eV] $  & 0.058318      & 0.069304                  & 0.058726 \\           
$m_{2}[eV] $  & -0.059139     & -0.07013                 & -0.058726 \\           
$m_{3}[eV] $  & -0.028399     & 0.0                      & 0.0 \\           
$s_{23} $     & 0.6745         & 0.6852                   & 0.6852 \\           
$s_{13}$      & 0.1329        & 0.0969                  & 0.0 \\           
$s_{12} $     & 0.7031        & 0.7050                  & 0.707107 \\             
$\bigtriangleup m_{12}^2[10^{-5}eV^2]$ & 9.64 & 11.52 & 0.0 \\         
$\bigtriangleup m_{23}^2[10^{-3}eV^2]$ & 2.69 & 4.92 & 3.45\\ \hline         
\end{tabular} 

\pagebreak

\vbox{
\noindent
\hfil
\vbox{
\epsfxsize=10cm
\epsffile [130 380 510 735] {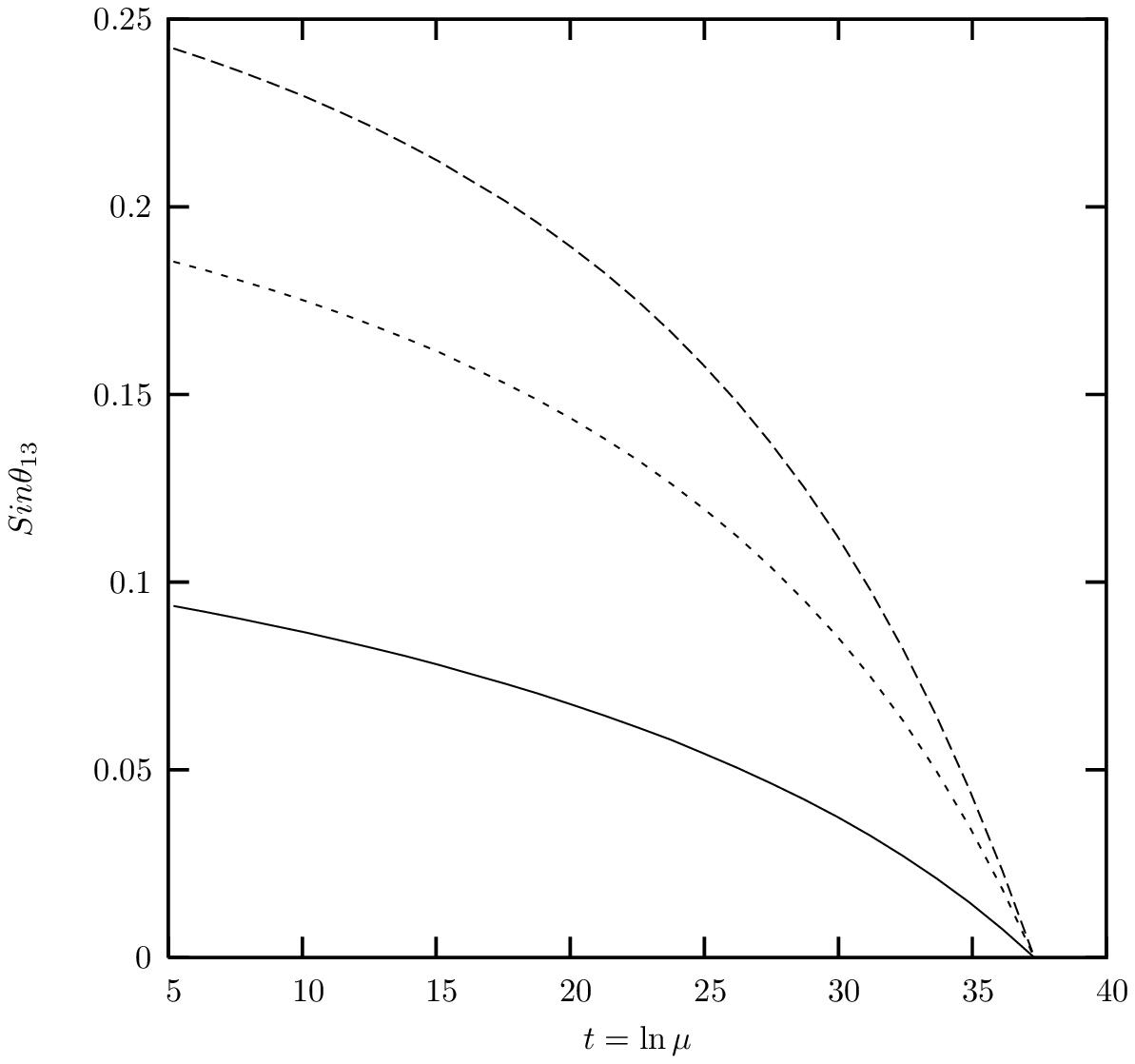}}
{\narrower\narrower\footnotesize\noindent
{Fig.1}
Radiative generation of the reactor  angle  $\sin\theta_{13}$ at low energy for three different sets of  input parameters
 at high scale, $M_R=1.6\times 10^{16} GeV$ as explained in the text ( Data taken from Tables 1 and 3).
\par}}
\vbox{
\noindent
\hfil
\vbox{
\epsfxsize=10cm
\epsffile [130 380 510 735] {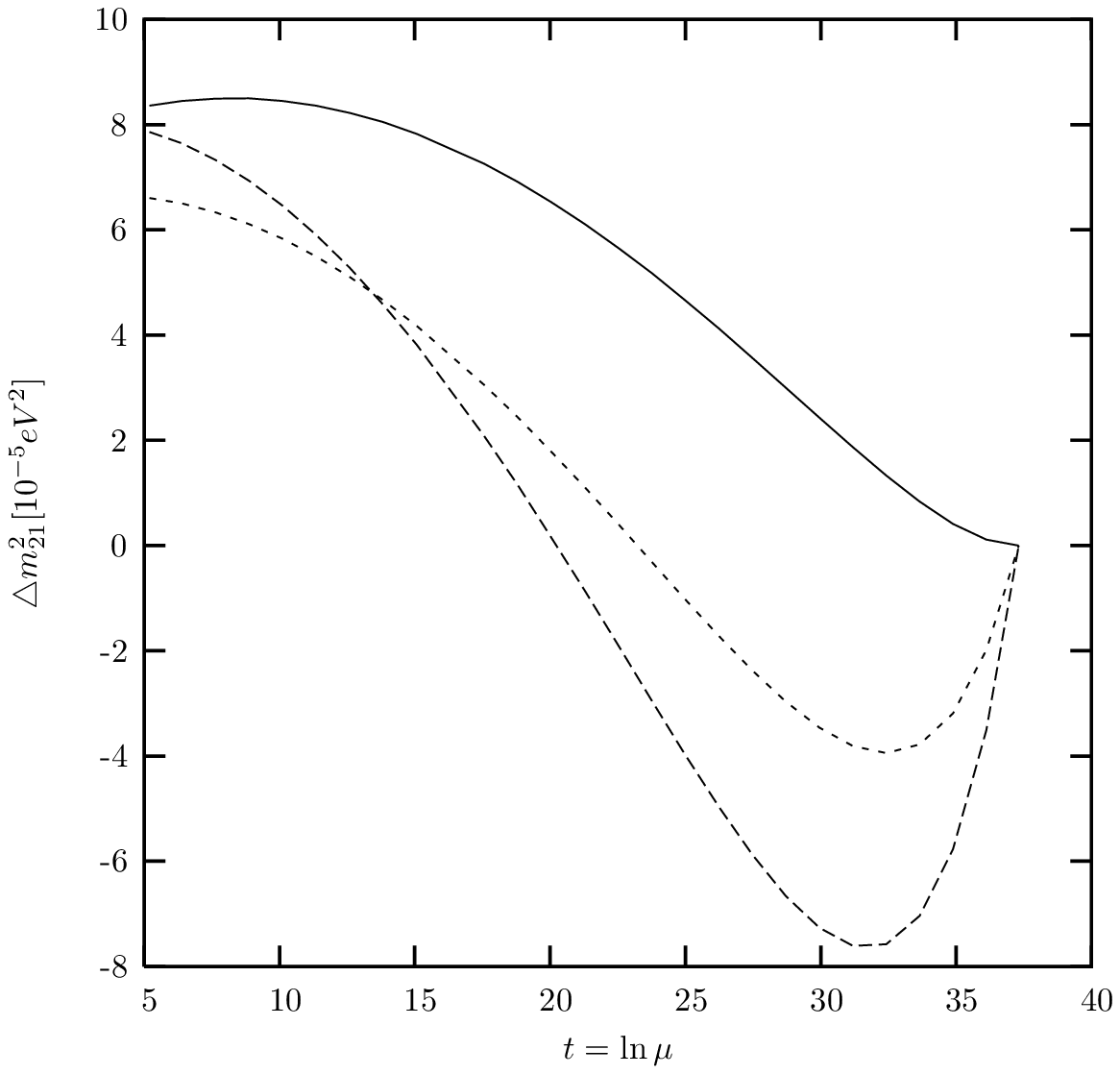}}
{\narrower\narrower\footnotesize\noindent
{Fig.2}
Radiative generation of solar mass scale $\bigtriangleup m^2_{21}$ at low energy for three different sets of  input parameters
 at high scale, $M_R=1.6\times 10^{16} GeV$ as explained in the text (Data taken from Tables 1 and 3). 
\par}}


\section{Summary and Discussion}
To summarise the main points in this paper, we first study the general RGEs for neutrino mass eigenvalues and mixing angles in MSSM 
and apply it to a model with  two degenerate  neutrinos having opposite CP parity. This leads to the JM conjecture [11] which means that 
 a non-zero reactor angle at high energy scale in the above two-fold degenerate model, 
can induce through radiative corrections the solar mass scale corresponding to LMA solution in MSSM.
We further show that  the expressions in JM conjecture [11]  are valid for static mixing angles and therefore, have limited predictions.
 We make a generalisation of JM conjecture by considering a  simultaneous 
running of neutrino masses and mixing angles and show that the above two-fold degenerate case  $m_i=(m, -m, m')$ with $m'\neq 0$ and $U_{e3}=0$ can
 also lead to radiative generation of non-zero values of $\bigtriangleup m^2_{21}$ and $U_{e3}$ at low scale in MSSM.    
This agrees with the results of our numerical analysis. In case if both $m'$ and $U_{e3}$ are zero at high scale, radiative generation of solar mass 
scale and CHOOZ mixing angle at low scale is not possible. Similar analysis with  an arbitrary CP violating phase will be interesting.  
 The present finding may have  important implications for model buildings from the  general framework of gauge symmetry. 

\section*{Acknowledgment}
N.N.S. thanks the High Energy Physics  Group, International Centre for Theoretical Physics, Trieste, Italy, 
for kind local hospitality during the course of the work.


\end{document}